\pdfoutput=1
\documentclass[11pt,twoside,a4paper,cmspaper,final,collab]{cms-tdr}
\def\svnVersion{89d32a0}\def\svnDate{2024/06/21}\def\cmsCernNoTag{CERN-EP-2023-146}\def\cmsCernDate{\today}\def\cmsMessage{Published in Physics Letters B as \href{http://dx.doi.org/10.1016/j.physletb.2024.138815}{\doi{10.1016/j.physletb.2024.138815}.}}
\begin{document}\cmsNoteHeader{TOP-22-008}

\newlength\cmsTabSkip\setlength{\cmsTabSkip}{1ex}
\ifthenelse{\boolean{cms@external}}{\providecommand{\cmsMiddle}{middle\xspace}}{\providecommand{\cmsMiddle}{middle\xspace}}
\ifthenelse{\boolean{cms@external}}{\providecommand{\cmsLeft}{upper\xspace}}{\providecommand{\cmsLeft}{left\xspace}}
\ifthenelse{\boolean{cms@external}}{\providecommand{\cmsRight}{lower\xspace}}{\providecommand{\cmsRight}{right\xspace}}
\ifthenelse{\boolean{cms@external}}{\providecommand{\cmsTable}[1]{\resizebox{\linewidth}{!}{#1}}}{\providecommand{\cmsTable}[1]{#1}}

\newcommand{\tWZ}{\ensuremath{\PQt\PW\PZ}\xspace}
\newcommand{\tZq}{\ensuremath{\PQt\PZ\PQq}\xspace}
\newcommand{\ttZ}{\ensuremath{\ttbar\PZ}\xspace}
\newcommand{\ttg}{\ensuremath{\ttbar\PGg}\xspace}
\newcommand{\ttW}{\ensuremath{\ttbar\PW}\xspace}
\newcommand{\ttVV}{\ensuremath{\ttbar\PV\PV}\xspace}
\newcommand{\ZZ}{\ensuremath{\PZ\PZ}\xspace}
\newcommand{\WZ}{\ensuremath{\PW\PZ}\xspace}
\newcommand{\WW}{\ensuremath{\PW\PW}\xspace}
\newcommand{\WZb}{\ensuremath{\PW\PZ\text{+}\PQb}\xspace}
\newcommand{\WZj}{\ensuremath{\PW\PZ\text{+jets}}\xspace}
\newcommand{\VVV}{\ensuremath{\PV\PV\PV}\xspace}
\newcommand{\VVmV}{\ensuremath{\PV\PV(\PV)}\xspace}
\newcommand{\Vg}{\ensuremath{\PV\PGg}\xspace}
\newcommand{\Xg}{\ensuremath{\PX\PGg}\xspace}
\newcommand{\tX}{\ensuremath{\PQt(\PAQt)\PX}\xspace}
\newcommand{\tWZxsecobs}{\ensuremath{354\pm54\stat\pm95\syst\unit{fb}}\xspace}
\newcommand{\tWZmuobs}{\ensuremath{2.6\pm0.4\stat\pm0.7\syst}\xspace}

\newcommand{\DPhi}{\ensuremath{\Delta\phi}\xspace}
\newcommand{\DEta}{\ensuremath{\Delta\eta}\xspace}
\newcommand{\delr}{\ensuremath{\DR}\xspace}

\newcommand{\DeepJet}{\textsc{DeepJet}\xspace}
\newcommand{\NNPDFThreeOne}{NNPDF3.1\xspace}

\newcommand{\PQj}{{\HepParticle{j}{}{}}\xspace}
\newcommand{\SRa}{\ensuremath{\text{SR}_{3\Pell,3\PQj}}\xspace}
\newcommand{\SRb}{\ensuremath{\text{SR}_{3\Pell,2\PQj}}\xspace}
\newcommand{\SRc}{\ensuremath{\text{SR}_{4\Pell}}\xspace}
\newcommand{\CRZZ}{\ensuremath{\text{CR}_{\ZZ}}\xspace}
\newcommand{\CRWZ}{\ensuremath{\text{CR}_{\WZ}}\xspace}
\newcommand{\SRBH}{\ensuremath{\text{SR}^{\text{Boosted}}_{\text{Had}}}\xspace}
\newcommand{\SRBL}{\ensuremath{\text{SR}^{\text{Boosted}}_{\text{Lep}}}\xspace}
\newcommand{\SRB}{\ensuremath{\text{SR}^{\text{Boosted}}}\xspace}
\newcommand{\msys}{\ensuremath{m_{\text{sys}}}\xspace}
\newcommand{\ptljmax}{\ensuremath{{\pt}_{\Pell\PQj}^{\text{max}}}\xspace}
\newcommand{\chithad}{\ensuremath{\chi^2_{\PQt,\text{had}}}\xspace}
\newcommand{\chitlep}{\ensuremath{\chi^2_{\PQt,\text{lep}}}\xspace}
\newcommand{\mthreej}{\ensuremath{m_{3\PQj}}\xspace}
\newcommand{\mt}{\ensuremath{m_\PQt}\xspace}
\newcommand{\thadres}{\ensuremath{\sigma_{\PQt,\text{had}}}\xspace}
\newcommand{\mlnj}{\ensuremath{m_{\Pell\PGn\PQj}}\xspace}
\newcommand{\tlepres}{\ensuremath{\sigma_{\PQt,\text{lep}}}\xspace}
\newcommand{\mtwoj}{\ensuremath{m_{\PQj\PQj}}\xspace}
\newcommand{\wres}{\ensuremath{\sigma_{\PW,\text{had}}}\xspace}

\newcommand{\mW}{\ensuremath{m_{\PW}}\xspace}
\newcommand{\mZ}{\ensuremath{m_{\PZ}}\xspace}
\newcommand{\mtop}{\ensuremath{m_{\PQt}}\xspace}

\cmsNoteHeader{TOP-22-008}
\title{Evidence for \tWZ production in proton-proton collisions at \texorpdfstring{$\sqrt{s} = 13\TeV$}{sqrt(s) = 13 TeV} in multilepton final states}

\date{\today}

\abstract{The first evidence for the standard model production of a top quark in association with a \PW boson and a \PZ boson is reported. The measurement is performed in multilepton final states, where the \PZ boson is reconstructed via its decays to electron or muon pairs and the \PW boson decays either to leptons or hadrons. The analysed data were recorded by the CMS experiment at the CERN LHC in 2016--2018 in proton-proton collisions at $\sqrt{s}=13\TeV$, and correspond to an integrated luminosity of 138\fbinv. The measured cross section is $354\pm 54\stat\pm 95\syst\unit{fb}$, and corresponds to a statistical significance of 3.4 standard deviations.}

\hypersetup{
pdfauthor={CMS Collaboration},
pdftitle={Evidence for tWZ production in proton-proton collisions at sqrt{s} = 13 TeV in multilepton final states},
pdfsubject={CMS},
pdfkeywords={CMS, single top quark, Z boson, machine learning, multilepton}}

\maketitle

\section{Introduction}
Measurements of rare processes are important at the CERN LHC, as these processes are tests of the validity of the standard model (SM) of particle physics, particularly in unexplored regions of the kinematic phase space. The electroweak production of the top quark in association with a \PW boson and a \PZ boson, \ie, the \tWZ process, has unique features that make it suitable to probe several interactions in the SM~\cite{Maltoni:2019aot,Faham:2021zet}. Its production cross section is calculated to be $136^{+9}_{-8}\unit{fb}$~\cite{Alwall:2014hca, Frederix2012} at next-to-leading order (NLO) in quantum chromodynamics (QCD). The main background process to \tWZ is the production of a top quark pair in association with a \PZ boson (\ttZ), which has a cross section of $859^{+76}_{-84}\unit{fb}$ at NLO (QCD and electroweak) and next-to-next-to-leading logarithmic accuracy~\cite{Kulesza:2020nfh}. Some diagrams of this process exhibit interference with the \tWZ process at NLO in QCD. The production of the two processes differs in principle in the number of resonant top quarks, which results in an additional jet originating from a \PQb quark (\PQb quark jet) in the final state of \ttZ, thus posing experimental challenges for the \tWZ measurement.

\begin{figure}[!ht]
\centering
\includegraphics[width=0.32\textwidth]{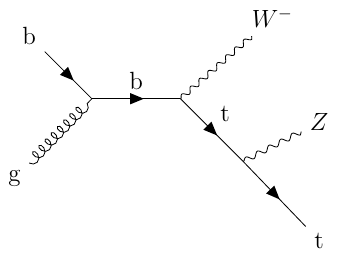}%
\hfill%
\includegraphics[width=0.32\textwidth]{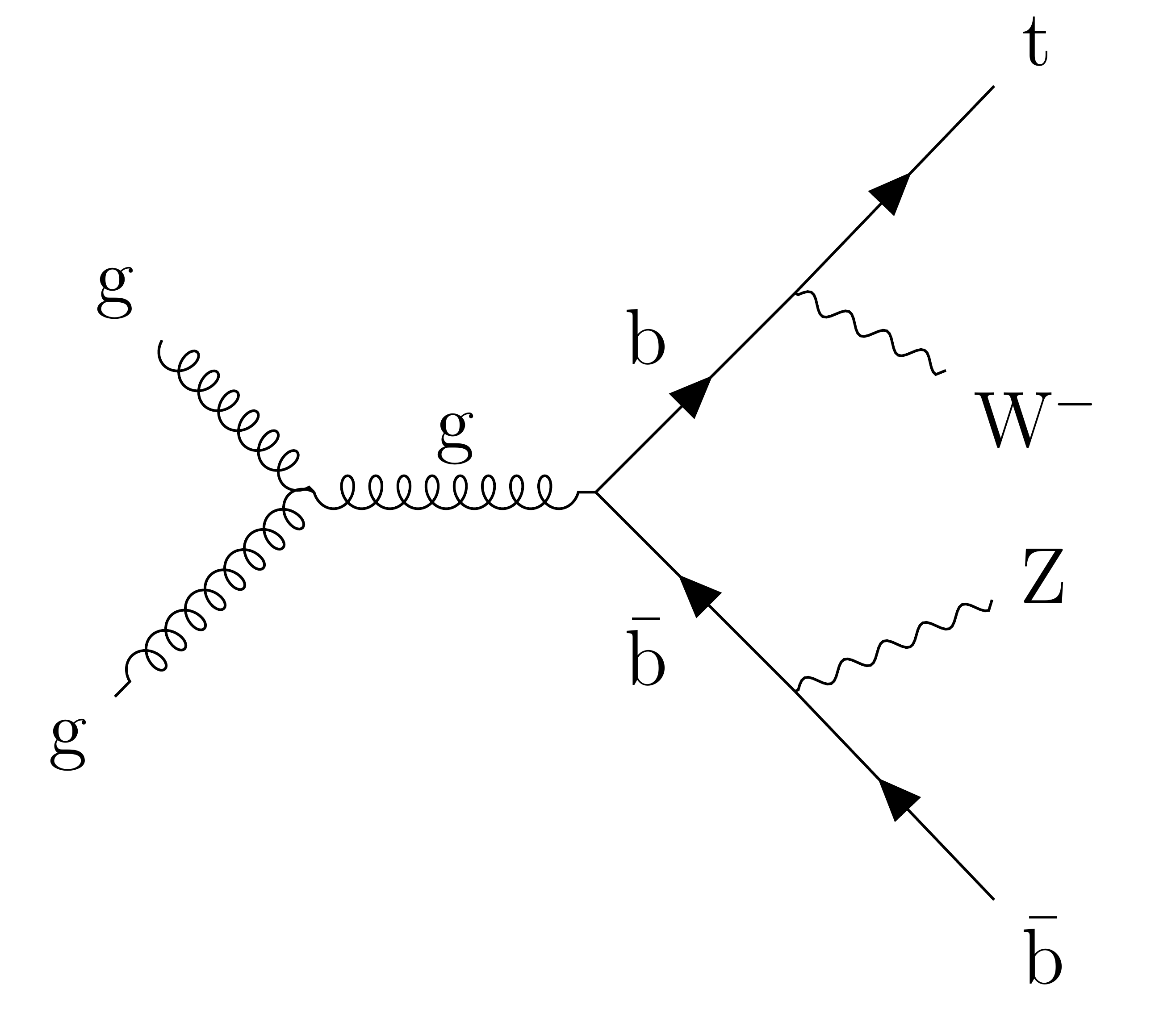}%
\hfill%
\includegraphics[width=0.32\textwidth]{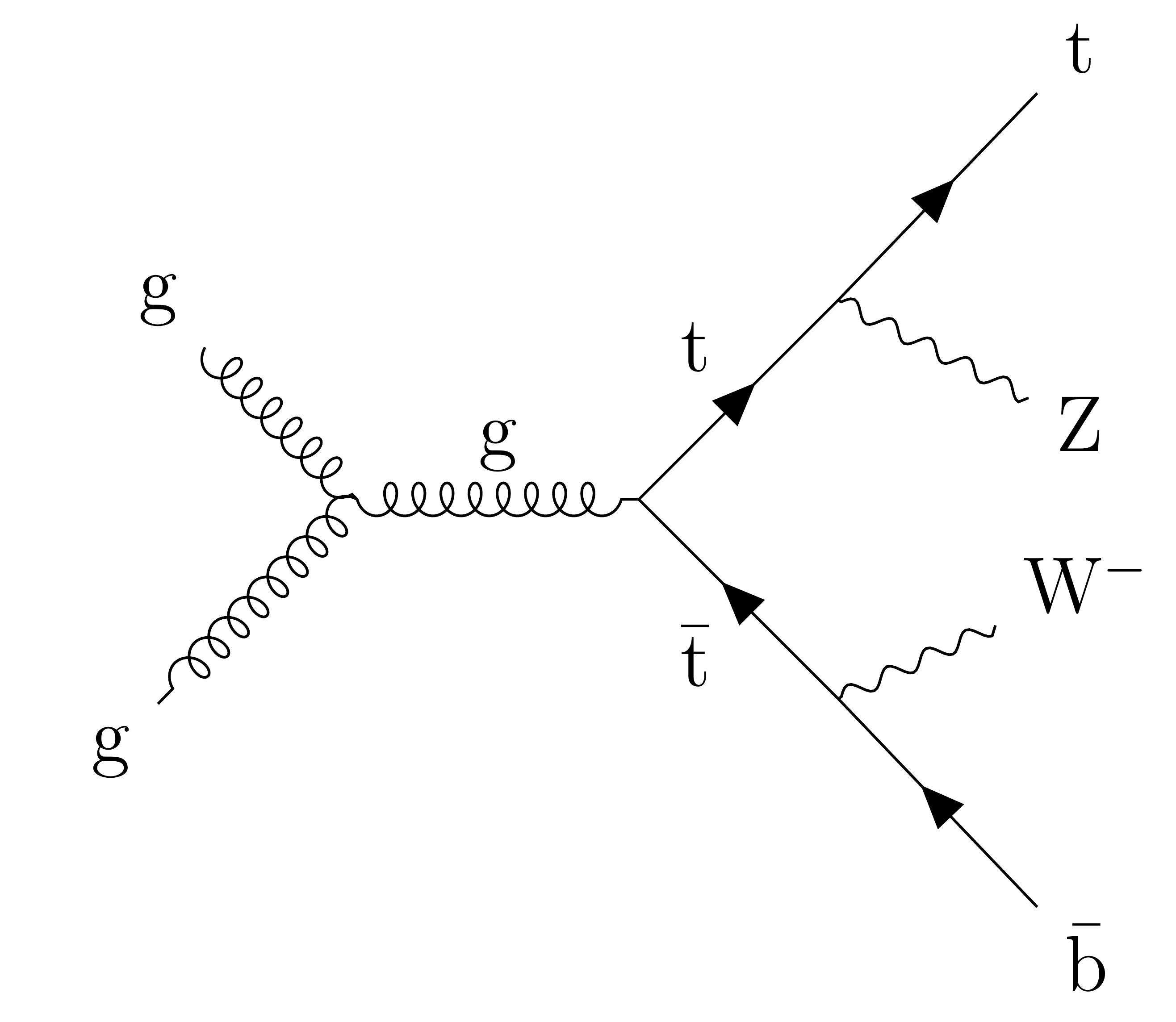}%
\caption{Feynman diagrams for the \tWZ production at LO (\cmsLeft) and NLO (\cmsMiddle) accuracy in QCD, and for the \ttZ production at the lowest order in perturbative QCD (\cmsRight). The \ttZ diagram can also arise when simulating the \tWZ process at NLO accuracy in QCD. The \cmsMiddle and \cmsRight diagrams share the same final state, and are different in the number of resonant top quarks.}
\label{fig:diagrams}
\end{figure}

This Letter reports the first experimental evidence for the \tWZ process. The measurement is performed using events with multiple leptons (electrons or muons), where a higher signal purity is expected compared to final states with hadronically decaying \PW or \PZ bosons. The top quark decays almost exclusively to a \PQb quark and a \PW boson. In the multilepton final state, the \PZ boson and at least one of the two \PW bosons decay to leptons, yielding a final state with three or four leptons.
Examples of Feynman diagrams for the \tWZ production at leading order (LO) and NLO accuracy in QCD are, respectively, shown in the \cmsLeft and \cmsMiddle parts of Fig.~\ref{fig:diagrams}. In Fig.~\ref{fig:diagrams} (\cmsRight), a Feynman diagram of the most challenging background is shown, yielding the same final state as that of the signal.

The analysed data set was recorded in proton-proton collisions at a center-of-mass energy of 13\TeV, collected by the CMS experiment at the LHC in 2016--2018, and corresponds to an integrated luminosity of 138\fbinv. This measurement approaches several unique challenges, from the simulation of the signal, where the interference with resonant \ttZ diagrams needs to be accounted for, to the use of multivariate analysis (MVA) techniques in order to achieve a more powerful discrimination between the signal and various background processes.
Moreover, the analysis is carried out in two different regions of the phase space, characterized by either the top quark being almost at rest, or having a large transverse momentum \pt, referred to as low- and high-\pt (or boosted) region, respectively. The low-\pt (boosted) region contains all events with a top quark $\pt \mathbin{< (>)} 270\GeV$. The addition of the boosted region brings an enhanced sensitivity to new phenomena in the \tWZ process, for instance in the context of SM effective field theory~\cite{Faham:2021zet,Keaveney:2021dfa}. In this Letter, we provide the inclusive measurement of the SM \tWZ cross section, setting the scene for the forthcoming analyses aimed at exploring the physics beyond the SM in this signature.

Tabulated results are provided in the HEPData record for this analysis~\cite{hepdata}.

\section{The CMS detector}
The central feature of the CMS apparatus is a superconducting solenoid of 6\unit{m} internal diameter, providing a magnetic field of 3.8\unit{T}. Within the solenoid volume are a silicon pixel and strip tracker, a lead tungstate crystal electromagnetic calorimeter (ECAL), and a brass and scintillator hadron calorimeter (HCAL), each composed of a barrel and two endcap sections. Forward calorimeters extend the pseudorapidity coverage provided by the barrel and endcap detectors. Muons are measured in gas-ionization detectors embedded in the steel flux-return yoke outside the solenoid. A more detailed description of the CMS detector, together with a definition of the coordinate system used and the relevant kinematic variables, can be found in Refs.~\cite{CMS:2008xjf, CMS:PRF-21-001}.

\section{Data and simulated event samples}
Events of interest are selected using a two-tiered trigger system. The first level (L1), composed of custom hardware processors, uses information from the calorimeters and muon detectors to select events at a rate of around 100\unit{kHz} within a fixed latency of 4\mus~\cite{CMS:2020cmk}. The second level, known as the high-level trigger (HLT), consists of a farm of processors running a version of the full event reconstruction software optimized for fast processing, and reduces the event rate to around 1\unit{kHz} before data storage~\cite{CMS:2016ngn}.
Collision events are selected using a combination of triggers that require the presence of either one, two, or three reconstructed leptons, muons or electrons, for further steps of the analysis. The trigger efficiency is close to 100\% for signal events passing the analysis selection~\cite{CMS:2021ugv}. 

Simulated Monte Carlo (MC) samples are produced with a consistent modeling of the detector operation conditions for each data-taking period. The signal sample is generated at NLO accuracy in QCD, in the five-flavor scheme (5FS), where bottom quarks are considered as sea quarks of the proton. The \MGvATNLO v2.6.5~\cite{Alwall:2014hca, Frederix2012} MC event generator is used for the hard scattering process, while parton shower (PS) and hadronization are modeled with \PYTHIA~v8.240~\cite{Sjostrand:2014zea}.
The parton distribution function (PDF) set is \NNPDFThreeOne~\cite{Ball:2017nwa} at next-to-NLO (NNLO) in 5FS, and the factorization and renormalization scales are configured dynamically.
At LO and in the 5FS, the \tWZ process can be easily identified through the tree-level partonic process $\Pg \PQb \to \tWZ$, and the simulation is straightforward. However, at NLO, real emissions of the type $\Pg \Pg \to \tWZ\PQb$ arise that may feature an additional resonant top \mbox{(anti-)quark} in the intermediate state. The process therefore overlaps with \ttZ and with \ttbar produced with an off-shell \PZ boson, at LO~\cite{Faham:2021zet}. A schematic illustration of a \ttZ diagram that appears in the generation of the signal sample at NLO in QCD is displayed in Fig.~\ref{fig:diagrams} (\cmsRight). Such overlap is removed by means of \textsc{MadSTR}~v1.0.0~\cite{Frixione:2019fxg}, which implements diagram removal (DR) and diagram subtraction (DS) schemes. In practice, two options are used for the generation of the samples: DR1, in which the squared resonant term from the squared amplitude of the process is removed together with the interference term between the resonant and nonresonant contributions; and DR2, where only the resonant term is removed. The DR1 scheme is used for the nominal samples, whereas DR2 is employed to assess systematic uncertainties in the signal model. The variation introduced by DS is not considered since it is consistently smaller than the one introduced by DR2~\cite{Demartin:2016axk}.

In order to increase the number of simulated events in the boosted region, more signal samples are generated with a top quark $\pt > 270\GeV$. To overcome the known issue that the DR and DS scheme are no longer effective at high momenta, an additional requirement on \PQb quarks is imposed at the generator level to retain events only from direct \tWZ production~\cite{Frixione:2019fxg,Faham:2021zet}. In practice, events with two \PQb quarks that originate from the top quark and have $\pt>30\GeV$ and pseudorapidity $\abs{\eta} < 2.4$ are removed, as they are likely to originate from the \ttZ process. The signal samples for low- and high-\pt top quarks are shown to have no overlapping events. All signal samples are generated with an on-shell \PZ boson decaying to charged leptons, and with at least one of the two \PW bosons decaying to a lepton and a neutrino. 

The \MGvATNLO MC event generator is used at NLO accuracy in QCD to model a number of background processes including \ttZ, the \ttbar production in association with a \PW boson and additional jets (\ttW), the single top quark production in association with a \PZ boson (\tZq), the associated production of a \PW boson and a \PZ boson (\WZ) and of two \PW bosons (\WW), the simultaneous production of three electroweak gauge bosons (\VVV, where \PV is either a \PW boson or a \PZ boson), and the production of a photon in association with a \PW or \PZ boson (\Vg). All the samples are generated using the 5FS, apart from the \tZq sample, generated in the 4FS, in which only up, down, charm, and strange quarks are considered as sea quarks of the proton. Other background processes, namely \PW boson production in association with jets, \ttbar production in association with a photon (\ttg), and with two electroweak gauge bosons (\ttVV) are simulated at LO accuracy with \MGvATNLO~\cite{Alwall:2014hca}. The \MCFM generator v.7.0.1~\cite{Campbell_2010} is used for the event simulation at LO for the gluon-initiated \ZZ production ($\Pg\Pg \to \ZZ$), while $\qqbar \to \ZZ$ and single top production in association with a \PW boson are simulated with \POWHEG~v2~\cite{Nason:2004rx, Frixione:2007vw, Alioli:2010xd, Hartanto_2015} at NLO accuracy in QCD. All events are processed with \PYTHIA\cite{Sjostrand:2014zea} to simulate PS and hadronization. The underlying event is modeled via the CP5 tune~\cite{TuneCP5} and the \NNPDFThreeOne at NNLO~\cite{Ball:2017nwa} set is used for the modeling of PDFs. The matching of the jets from matrix elements to those from PS is performed using the MLM scheme~\cite{Alwall_2007} for LO and the FxFx merging scheme~\cite{Frederix2012} for the NLO samples that are generated using \MGvATNLO. For all processes, the detector response is simulated using a detailed description of the CMS apparatus, based on the \GEANTfour toolkit~\cite{GEANT4}. Additional proton-proton interactions in the same and/or neighboring bunch crossings (pileup) are simulated with \PYTHIA8. All background processes are normalized according to the measured integrated luminosity and to their most accurate theoretical cross sections (see \eg Refs.~\cite{Kulesza:2020nfh,Czakon:2013goa,pagani2020nlo,Grazzini2020,Frederix2021}). An important source of background is the contribution of the so-called ``nonprompt'' leptons from hadron decays, or from misidentification of jets or hadrons in the reconstruction process, as opposed to the prompt leptons originating from the \PW and \PZ boson decays. The processes that mostly contribute with nonprompt leptons in this measurement are Drell--Yan and \ttbar in the dileptonic channel. This background cannot be accurately described by simulation and is estimated from data.

\section{Event reconstruction and selection}
The particle-flow algorithm~\cite{CMS:2017yfk} is used to reconstruct and identify individual particles in the event, combining information from different parts of the detector. The primary vertex (PV) is taken to be the vertex corresponding to the hardest scattering in the event, identified using the tracking information alone, as described in Section 9.4.1 of Ref.~\cite{CMS-TDR-15-02}. The energy of electrons is determined from a combination of the electron momentum at the PV, the energy of the corresponding ECAL cluster, and the energy sum of all bremsstrahlung photons spatially compatible with originating from the electron track. The energy of muons is obtained from the curvature of the corresponding track. The energy of charged hadrons is evaluated via a combination of their momentum measured in the tracker and the matching of the ECAL and HCAL energy deposits, corrected for the response function of the calorimeters to hadronic showers. Finally, the energy of neutral hadrons is obtained from the corresponding corrected ECAL and HCAL energies.

Reconstructed electrons (muons) are required to have $\pt > 10\GeV$ and $\abs{\eta} < 2.5$ (2.4). In order to increase the purity of prompt leptons, an MVA-based discriminant is used~\cite{CMS:2023ftu,CMS-PAS-MUO-22-001} to identify the leptons that enter the final selection. Leptons that fail this requirement are subjected to additional criteria and used in the estimation of the nonprompt-lepton contribution from control regions (CRs) in data.

Jets are reconstructed by clustering the particle-flow candidates using the anti-\kt algorithm~\cite{Cacciari:2008gp, Cacciari:2011ma} with a distance parameter of 0.4. They are required to have $\pt > 25\GeV$, $\abs{\eta} < 2.5$, and to be separated from selected leptons by a distance $\delr = \sqrt{\smash[b]{(\DEta)^2 + (\DPhi)^2}} > 0.4$. Here, \DEta and \DPhi are the separation in pseudorapidity and azimuthal angle between the jet and the lepton, respectively. The products of hadronically decaying top quarks with \pt more than about twice the top quark mass, $\mtop=172.5\GeV$, are collimated and identified as a single large-$R$ jet. Such jets are reconstructed using the anti-\kt algorithm with a distance parameter of 0.8, and are employed in the analysis of the boosted region. They are required to have $\pt > 300\GeV$ and $\abs{\eta} < 2.4$. The magnitude of the missing transverse momentum vector \ptvecmiss, defined as the projection onto the transverse plane relative to the beam axis, of the negative vector sum of the momenta of all particle-flow candidates, is referred to as \ptmiss~\cite{metPF}. Corrections to the jet energies are propagated to \ptvecmiss.

The \DeepJet algorithm~\cite{BTV-16-002, Bols:2020bkb, CMS:DP-2023-005} is used to identify \PQb quark jets. Jets are considered \PQb tagged if they pass a requirement on the \DeepJet score that provides a misidentification rate of 1\% for jets originating from \PQu, \PQd, or \PQs quarks and gluons. Jets that are \PQb tagged are referred to as ``\PQb jets'' in the following.

The signal region (SR) is first divided into two parts depending on the transverse momentum of the top quark.
Events in the low-\pt region are required to have two leptons with $\pt>20\GeV$ and one with $\pt>25\GeV$. Two of the leptons must have opposite electric charge and same flavor (OCSF). The invariant mass of the OCSF lepton pair must lie within $\mZ\pm 15\GeV$, where \mZ is the \PZ boson mass equal to 91.2\GeV~\cite{EXT:PDG-2022}. Selected events should contain at least two jets of which at least one is \PQb tagged. To increase the sensitivity of the analysis, the low-\pt region is further split into \SRa and \SRb categories. Events with at least three jets fall into the \SRa category whereas \SRb contains events with exactly two jets. The \SRa category has the largest contribution of signal with 77 signal events expected over 1249 background events, of which more than half are expected to be \ttZ. The expected signal yield in the \SRb category is 28 to compare with 822 background events. Background in this category is mostly from nonprompt leptons and from \WZ in association with light-flavor or gluon jets (\WZj). The \ttZ process contributes to less than 15\% of the background in \SRb. Another category in the low-\pt region targets the fully-leptonic decay of the \tWZ signal. Events in this category, \SRc, must have at least one \PQb-tagged jet and four leptons, three of which respect the same \pt requirement of \SRa and a fourth one with $\pt>10\GeV$. While two out of four leptons are required to be OCSF, with an invariant mass within $\mZ\pm 15\GeV$, the other two are required to fail at least one of these criteria. In this region, 16 signal events are expected over 160 background events, of which more than 70\% is \ttZ.

The high-\pt region is divided in two categories, \SRBH and \SRBL, respectively targeting hadronically and leptonically decaying top quarks. The selection in both categories follows that of \SRa with a number of additional requirements.
The \SRBH category is required to contain a candidate for the boosted top quark, a large-$R$ jet, with a soft-drop~\cite{Larkoski:2014wba} mass between 105 and 210 GeV, in the vicinity of $\delr=0.8$ around a \PQb-tagged jet. 
In the \SRBL category, the lepton that is not compatible with the hypothesis of originating from the \PZ boson (the third lepton) is required to have $\pt > 30\GeV$, and to lie in a cone of $\delr=2$ around a \PQb-tagged jet with $\pt > 200\GeV$. A fully connected deep neural network (DNN)~\cite{krizhevsky2012imagenet} is trained on simulated \ttbar events, with one top quark decaying into a lepton, with the aim of tagging the lepton from the high-\pt top quark decay, using TensorFlow v2.8.0~\cite{tensorflow2015}. Each third lepton is assigned a score according to its compatibility with the hypothesis of originating from the high-\pt top quark. Variables related to the third lepton, the \PQb jet, as well as a combination of the two, are fed into the DNN, which is then validated on simulated \ttbar events with both top quarks decaying into leptons. Overall in the high-\pt region, 5 signal and 81 background events are expected.

In addition to regions enriched in signal events, two regions are defined in data to control and validate the \ZZ and \WZ backgrounds, in order to ensure a correct modelling. Events containing exactly four leptons, with the two pairs being compatible with the hypothesis of originating from a \PZ boson, are collected in the region enriched in \ZZ, denoted by \CRZZ. The region enriched in \WZj events, \CRWZ, is built with events containing exactly three leptons, two of them originating from the \PZ boson, with $\ptmiss > 50\GeV$, and no \PQb-tagged jets. The expected yields for the signal and background processes, together with the observed number of events in each region, are reported in Table~\ref{tab:yields1} and Table~\ref{tab:yields2}.

\begin{table*}[!ht]
\topcaption{Expected yields for signal and background processes and observed number of events in the signal regions. Uncertainties arise from the limited size of the simulated samples, except for the nonprompt-lepton contribution which is extracted from data and assigned a 30\% uncertainty.}
\centering
\renewcommand{\arraystretch}{1.2}
\begin{tabular}{lcccccc}
    \hline
    {\LARGE\strut} & \SRa & \SRb & \SRc & \SRB \\
    \hline
    \tWZ signal & $ 77.47 \pm 0.12 $ & $ 28.19 \pm 0.07 $ & $ 15.98 \pm 0.06  $ & $5.44\pm 0.02$ \\[\cmsTabSkip]
    \ttZ &$ 657.9 \pm 1.6 $ &$ 122.76 \pm 0.61 $ &$ 113.86 \pm 0.64 $ &$ 59.03 \pm 0.50$ \\
    Nonprompt leptons &$ 139 \pm 42 $ &$ 170 \pm 51 $ &$ 1.02 \pm 0.31 $ &$ 1.94 \pm 0.58$ \\
    \tZq &$ 86.45 \pm 0.78 $ &$ 108.69 \pm 0.71 $ &$ 0.29 \pm 0.04 $ &$ 4.37 \pm 0.17$ \\
    \ZZ &$ 22.7 \pm 2.4 $ &$ 60.6 \pm 4.1 $ &$ 20.0 \pm 2.3 $ &$ 0.30 \pm 0.29$ \\
    \WZ &$ 166.4 \pm 3.3 $ &$ 227.8 \pm 4.0 $ &$ 0.59 \pm 0.19 $ &$ 6.84 \pm 0.66$ \\
    \VVmV &$ 15.51 \pm 0.11 $ &$ 10.55 \pm 0.09 $ &$ 1.35 \pm 0.03 $ &$ 0.64 \pm 0.02$ \\
    \tX &$ 108.30 \pm 0.99 $ &$ 49.4 \pm 1.2 $ &$ 17.32 \pm 0.34 $ &$ 6.26 \pm 0.19$ \\
    \Xg &$ 54.1 \pm 2.6 $ &$ 78.3 \pm 3.7 $ &$ 6.92  \pm 0.95 $ &$ 1.08 \pm 0.31$ \\[\cmsTabSkip]
    Total backgrounds &$ 1249 \pm 42 $ &$ 822  \pm 51 $ &$ 159.9 \pm 2.6 $ &$ 80.8 \pm 1.1$ \\[\cmsTabSkip]
    Data & 1463 & 849 & 180 & 77 \\
    \hline
\end{tabular}
\label{tab:yields1}
\end{table*}

\begin{table}[!ht]
    \topcaption{Expected yields for signal and background processes and observed number of events in the control regions. Uncertainties arise from the limited size of the simulated samples, except for the nonprompt-lepton contribution which is extracted from data and assigned a 30\% uncertainty.}
    \centering
\renewcommand{\arraystretch}{1.2}
\cmsTable{\begin{tabular}{lcccccc}
    \hline
    {\LARGE\strut} & \CRWZ & \CRZZ \\
    \hline
    \tWZ signal &  $31.96 \pm 0.08$ & $2.39 \pm 0.02$ \\[\cmsTabSkip]
    \ttZ & $112.41 \pm 0.73$ & $14.44 \pm 0.23$ \\
    Nonprompt leptons & $1450 \pm 430$ & $23.0 \pm 7.0$ \\
    \tZq & $74.84 \pm 0.67$ & $0.05 \pm 0.01$ \\
    \ZZ & $597 \pm 12$ & $2202 \pm 22$ \\
    \WZ & $10610 \pm 25$ & $0.68 \pm 0.16$ \\
    \VVmV & $166 \pm 14$ & $16.52 \pm 0.07$\\
    \tX & $39.4 \pm 1.8$ & $1.08 \pm 0.07$ \\
    \Xg & $519 \pm 11$ & $2.53 \pm 0.60$ \\[\cmsTabSkip]
    Total backgrounds & $13520 \pm 430$ & $2028 \pm 23$  \\[\cmsTabSkip]
    Data & 12743 & 2352 \\
    \hline
\end{tabular}}
\label{tab:yields2}
\end{table}

\section{Background estimation}
The contribution of nonprompt leptons is estimated from data in a similar way to that of Ref.~\cite{CMS:2021ugv}. Event weights, derived from simulated QCD multijet samples, are assigned to collision events that are selected similarly to signal with modified (looser) lepton requirements. To validate the method, a data control region enriched in the nonprompt contribution is defined by a trilepton selection similar to that of signal, except that the OCSF pair is required to be out of the \PZ boson mass window. Events should also have at least one \PQb-tagged jet. The selection provides a good amount of nonprompt leptons. Considering key distributions such as the \pt of the trailing lepton, the nonprompt estimation is found to be compatible with data within 30\%.

\section{Multivariate classifiers and statistical analysis}
\begin{figure*}[!ht]
\centering
\includegraphics[width=0.47\textwidth]{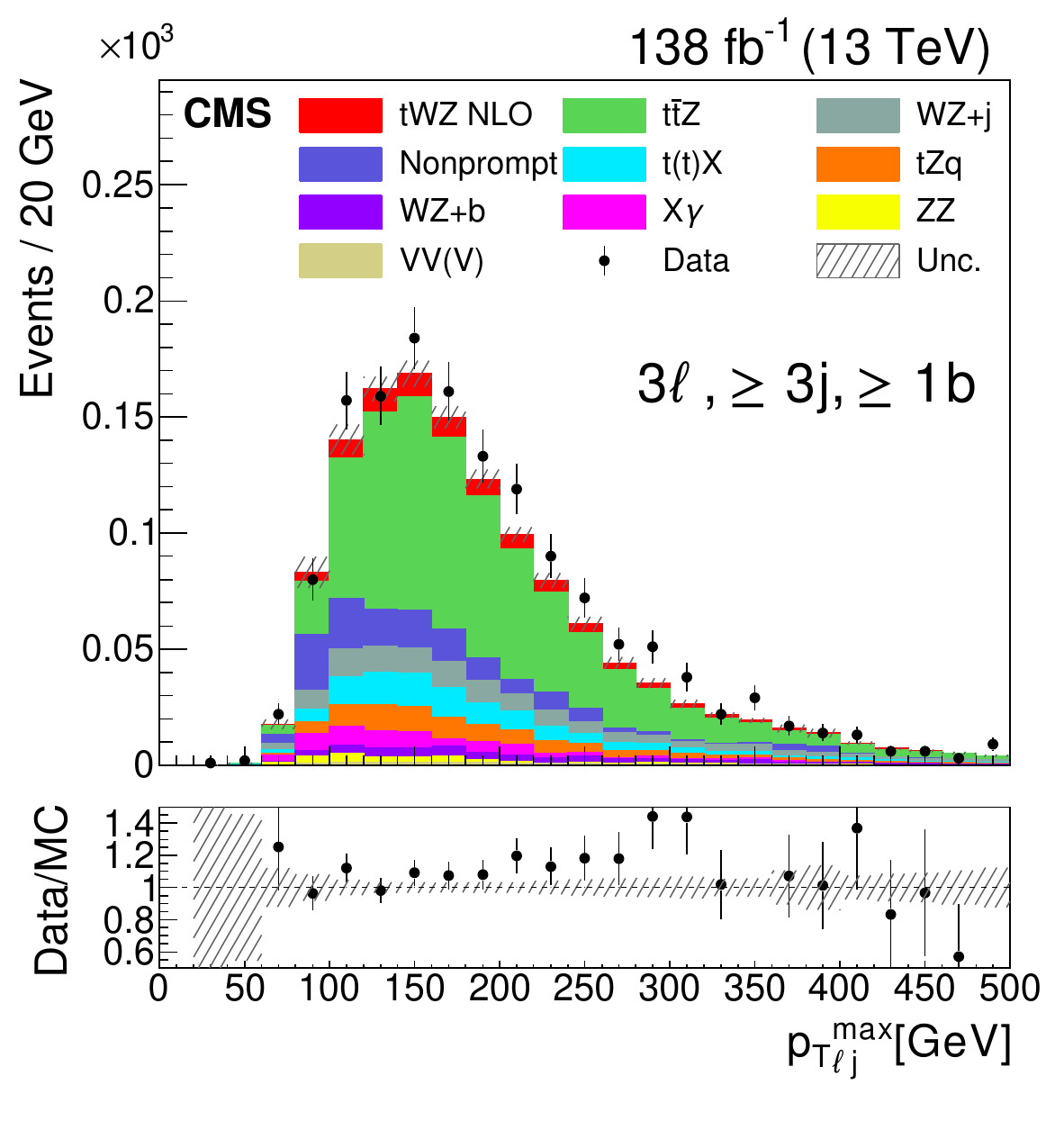}%
\hfill%
\includegraphics[width=0.47\textwidth]{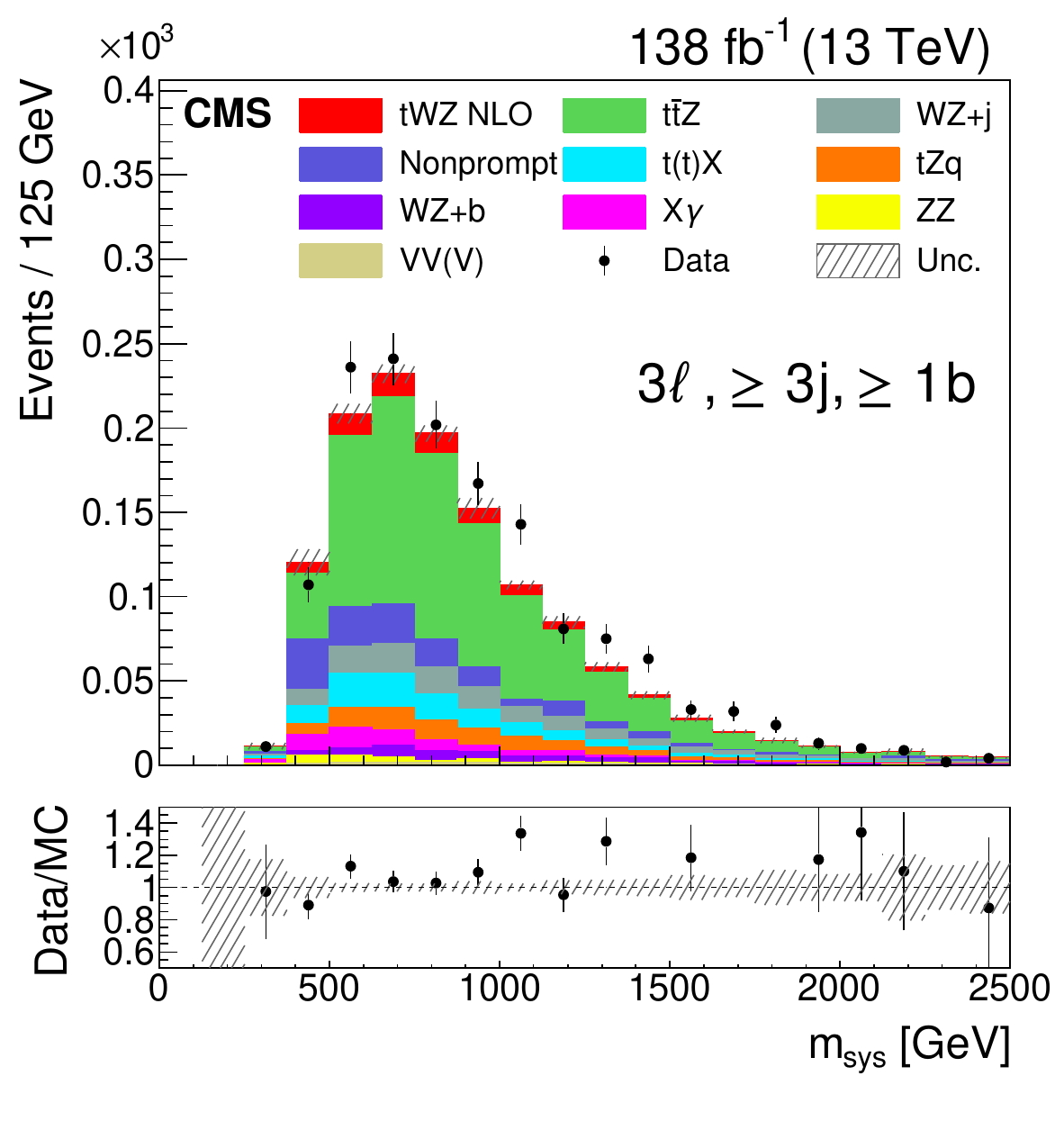}\\
\includegraphics[width=0.47\textwidth]{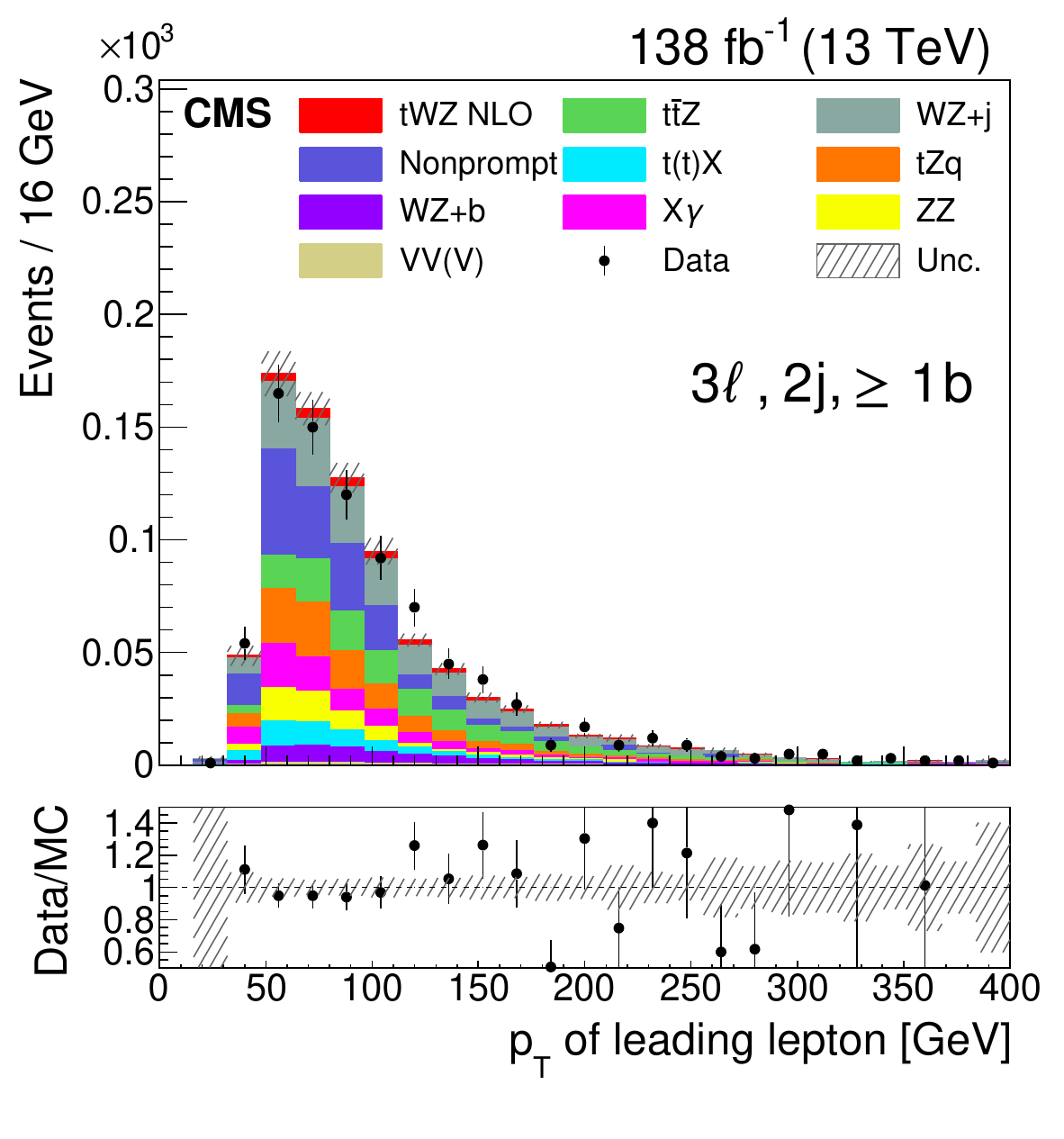}%
\hfill%
\includegraphics[width=0.47\textwidth]{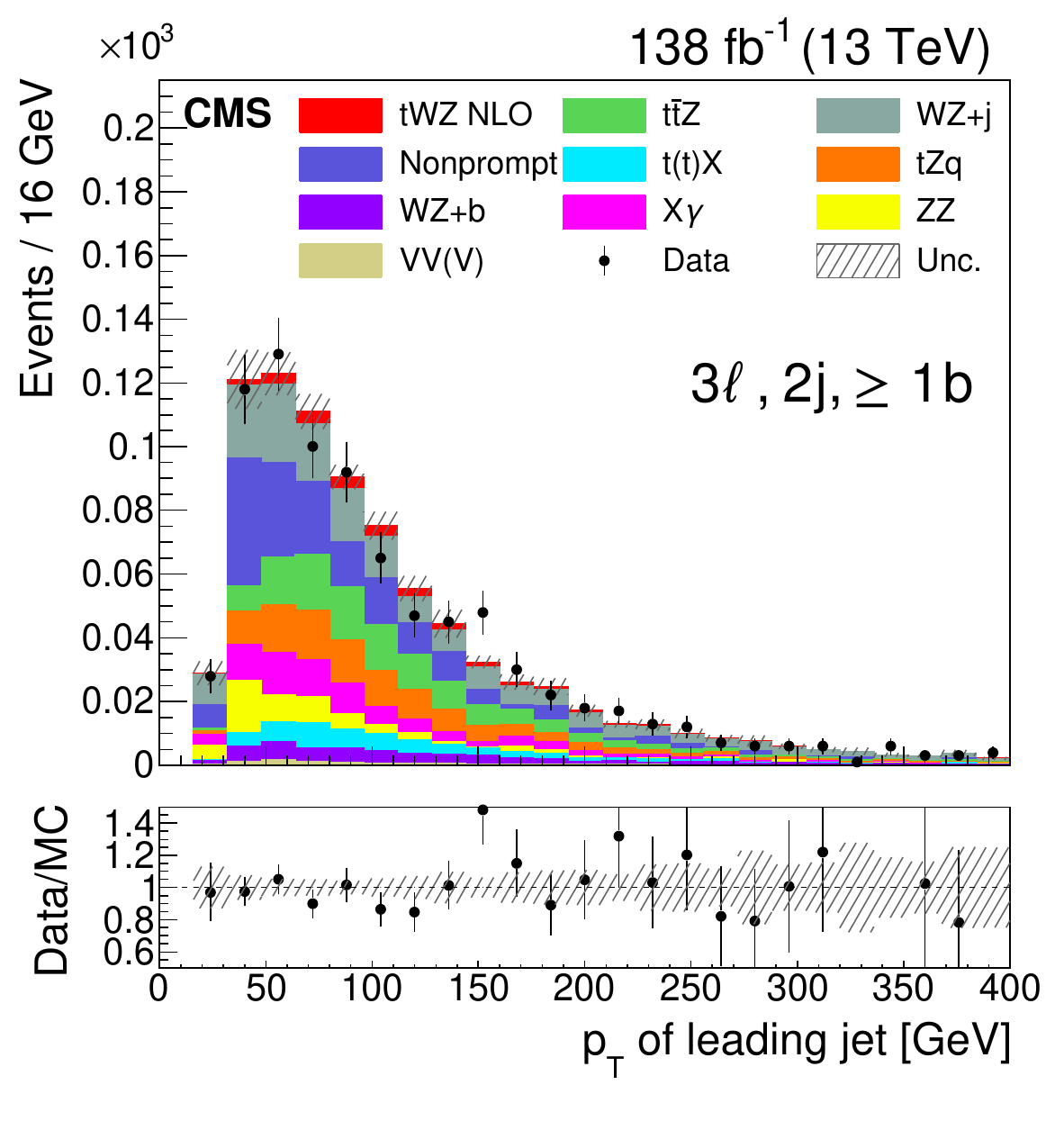}%
\caption{Distributions of some of the input features for the two DNN trainings in data, compared to the expected yield of signal and backgrounds: the \ptljmax (upper left) and \msys (upper right) observables in \SRa, as well as the \pt of the leading lepton (lower left)  and leading jet (lower right) in \SRb. The \VVmV group in the legend denotes the \VVV, \WW, and \PW in association with jets backgrounds. The dashed band shows the total uncertainty (statistical and systematic) before the fit.}
\label{fig:input}
\end{figure*}

In order to increase the discrimination power between signal and backgrounds, DNN classifiers are designed and implemented via TensorFlow in \SRa and \SRb.  A fully connected DNN is trained on simulated events in \SRa in the form of a multiclass classifier. Events are given a score according to their compatibility with \tWZ, \ttZ, and other background events, and are therefore classified in three categories. The following observables yield discrimination between the processes and are fed into the classifier: the number of \PQb jets, the \pt and $\eta$ of jets and leptons, the maximum \delr between lepton pair combinations, the maximum \delr and invariant mass between jet pair combinations, the maximum $\eta$ of all jets in the event, the maximum \pt of the lepton-jet combinations, \ptljmax, and the invariant mass of the system of leptons, jets, and \ptvecmiss, \msys. The last two variables are shown in Fig.~\ref{fig:input} (upper row).

To further distinguish the \tWZ signal from \ttZ production, the following kinematic observables specific to the top quark and \PW boson are added as inputs to the classifier: the mass, \pt, $\eta$, and $\phi$ of the top quark and hadronically decaying \PW boson, as well as the \pt, $\eta$, and $\phi$ of the leptonically decaying \PW boson. This requires the reconstruction of the signal event hypothesis, taking into account the origin of the third lepton. Two top quark decay hypotheses are checked per event using \chithad and \chitlep variables,
\begin{linenomath}\ifthenelse{\boolean{cms@external}}{\begin{multline}\label{eq:chi2}
    \chithad=\left( \dfrac{\mthreej - \mt}{\thadres} \right)^2
    \quad\text{and}\\
    \chitlep = \left( \dfrac{\mlnj - \mt}{\tlepres} \right)^2 + \left( \dfrac{\mtwoj - \mW}{\wres} \right)^2.
\end{multline}}{\begin{equation}\label{eq:chi2}
    \chithad=\left( \dfrac{\mthreej - \mt}{\thadres} \right)^2
    \quad\text{and}\quad
    \chitlep = \left( \dfrac{\mlnj - \mt}{\tlepres} \right)^2 + \left( \dfrac{\mtwoj - \mW}{\wres} \right)^2.
\end{equation}}\end{linenomath}
In Eq.~\eqref{eq:chi2}, \mthreej and \mtwoj are the tri- and di-jet invariant masses, respectively, expected to be compatible with the mass of a top quark and a \PW boson that decay hadronically. The invariant mass \mlnj of the lepton not coming from the \PZ boson, the neutrino, and a jet is expected to be close to the mass of the leptonically decaying top quark. The variables \thadres, \tlepres, and \wres represent the mass resolutions and are estimated from simulation. The \PW boson mass is set to $\mW=80.4\GeV$~\cite{EXT:PDG-2022} and the \mtop is set to 172.5\GeV, following the settings in the simulation. The neutrino four-momentum in the \PW boson decay is estimated as described in Ref.~\cite{Chatrchyan_2011}.
All lepton-jet combinations are considered and a set of values for \chitlep and \chithad are obtained. The combination corresponding to the smallest value is taken as the reconstructed event hypothesis, identifying also the top quark decay. The kinematic properties of the top quark and, for the leptonic top quark decays, of the hadronically decaying \PW boson are used in the DNN training.

Another DNN is trained as a binary classifier on simulated events in \SRb, with the aim of discriminating the signal from all other background sources, especially nonprompt leptons and \WZj. A broad range of observables are investigated, and those that provide the best discrimination between signal and background processes are selected: the \pt and $\eta$ of the leptons and jets, the number of \PQb jets, the \pt and invariant mass of the jet pair, the \msys, and the \pt of the system of leptons and jets. Figure~\ref{fig:input} (lower row) shows the \pt of the leading lepton and leading jet as examples of input features.

In order to avoid overtraining, both the multiclassifier used in \SRa and the binary classifier used in \SRb are trained in two copies, so that each version could be trained and used on complementary halves of the simulated samples. 

The \tWZ cross section is measured via the maximization of a binned likelihood function in a fit to data. This function combines the Poisson probabilities for the observed yields given the predicted signal and background estimates in each bin, and incorporates as nuisance parameters all sources of systematic uncertainties that may affect the number of observed signal or background events. The fit returns the measured value of the signal strength, defined by the ratio of the measured cross section and the expected one, together with its uncertainty.
The fit is performed using the CMS statistical analysis and combination tool \textsc{Combine}\xspace~\cite{CAT-23-001}.

\begin{figure*}[!ht]
\centering
\includegraphics[width=0.47\textwidth]{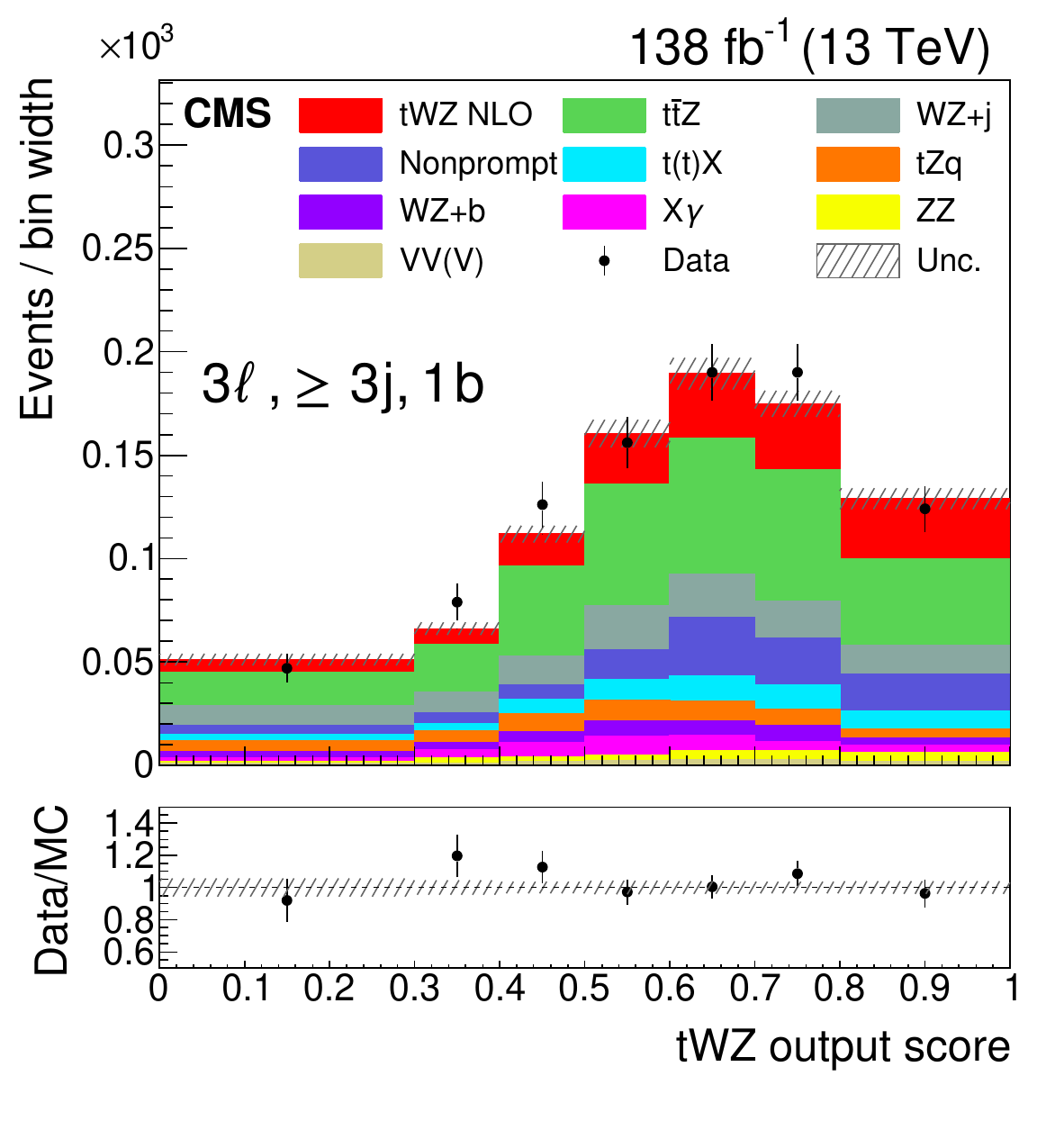}%
\hfill%
\includegraphics[width=0.47\textwidth]{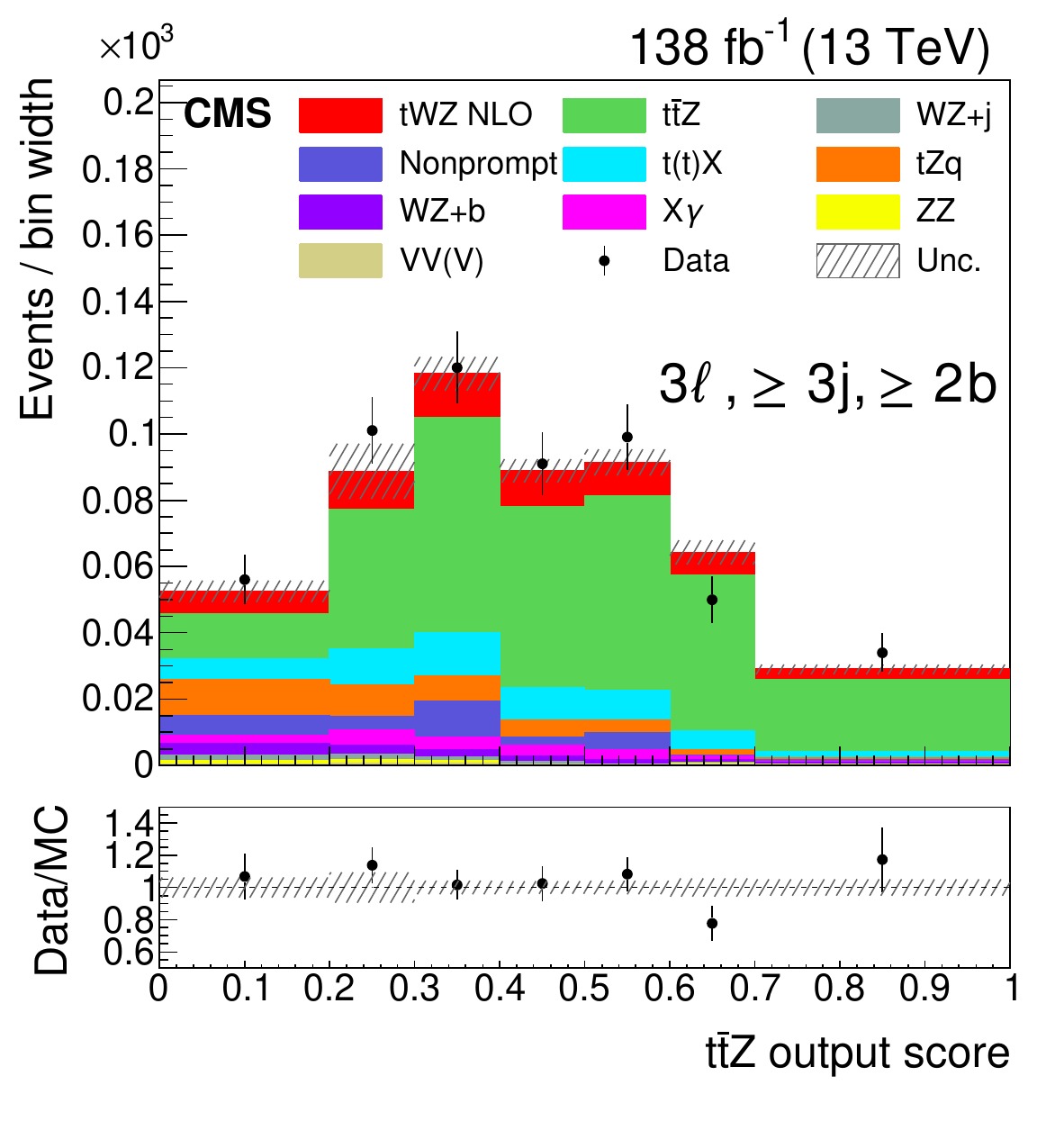}\\
\includegraphics[width=0.47\textwidth]{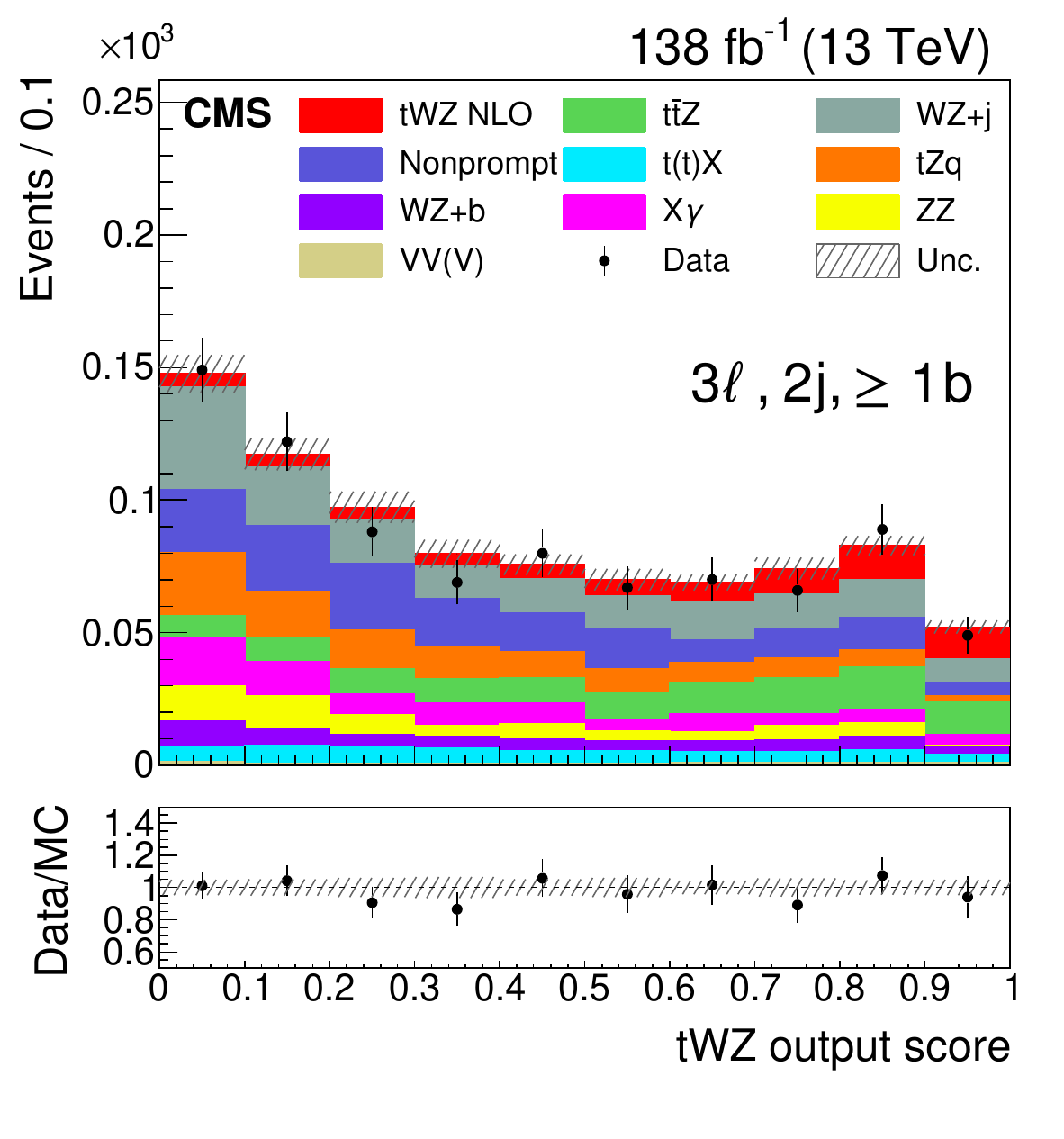}%
\hfill%
\includegraphics[width=0.47\textwidth]{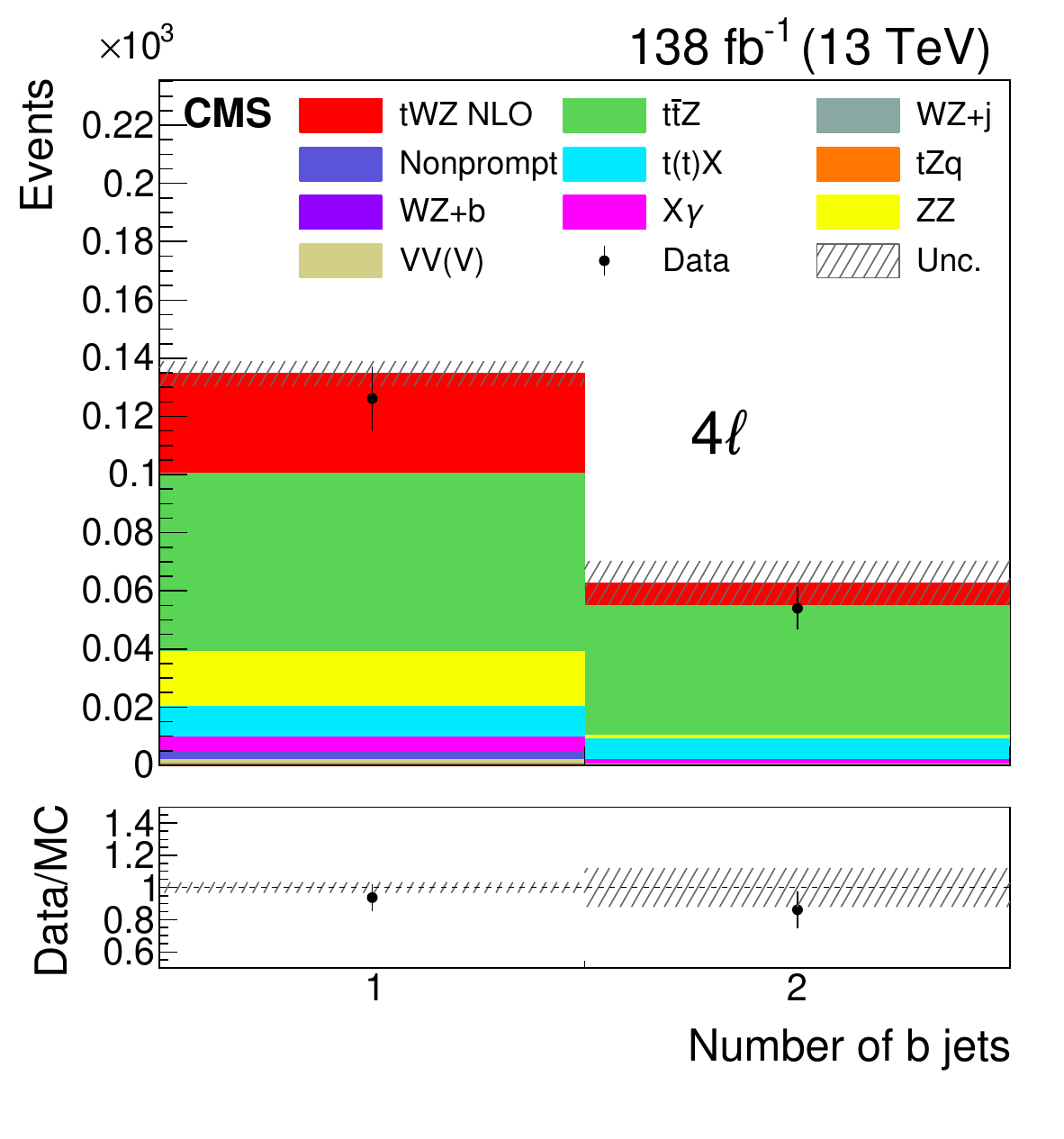}%
\caption{Score of the \tWZ output node from the multiclass classifier in \SRa for events with exactly one \PQb jet (upper left), and of the  \ttZ output node in \SRa for events with more than one \PQb jet (upper right);  score of the \tWZ output node of the binary classifier in \SRb (lower left), and the \PQb jet multiplicity in \SRc (lower right). The \VVmV group in the legend denotes the \VVV, \WW, and \PW in association with jets backgrounds. The dashed band shows the total uncertainty (statistical and systematic) after the fit.}
\label{fig:dnnscores}
\end{figure*}

Inputs to the binned likelihood function are distributions and number of events, depending on the region and category. The \SRa category is split into two parts: one containing events with exactly one \PQb jet where the distribution of the \tWZ output node score is taken from, and the other containing the rest and providing the distribution of the \ttZ output node score to the fit. This is to enhance the sensitivity of the fit to differences between \tWZ and \ttZ, considering the additional \PQb jet in the \ttZ background. From \SRb, the distribution of the \tWZ output node score contributes to the fit whereas the \PQb jet multiplicity distribution is taken from \SRc. Given the limited number of events in \SRBH and \SRBL, the two SRs are merged and contribute to the signal extraction fit via event counting. The number of events in  \CRZZ and \CRWZ are also included in the fit in order to constrain the normalization of the \ZZ and \WZj processes. To account for the different \PQb quark content in the SR and \CRWZ, the \WZ background is split into \WZb and \WZj based on the generator-level information.
The \tWZ and \ttZ output node scores from the multiclass classifier in \SRa and the score of the binary classifier in \SRb are shown in Fig.~\ref{fig:dnnscores} after the fit (upper and lower rows, respectively).

\section{Systematic uncertainties}
Several sources of systematic uncertainty are considered. We report the uncertainties in the \PQb tagging, lepton, and trigger efficiency corrections, the distribution of pileup events in simulation, the jet energy correction (JEC) and resolution (JER), and the unclustered missing energy. The uncertainties arising from our imperfect knowledge encoded in the MC event generator parameters and settings are also included in the fit. They include systematic effects associated with the matrix element renormalization and factorization scales, the PS simulation, and the PDFs (the latter being estimated by reweighting the simulation using the corresponding variations in the {NNPDF} sets). Furthermore, variations in templates of the \tWZ signal arising from the difference between the DR1 and DR2 schemes are included as uncertainties associated with the signal modeling. Theoretical uncertainties in the cross sections of the background processes are also taken into account. In particular, a normalization uncertainty of 15\% is estimated for the \ttZ cross section, which covers the relative difference between the measurement of the cross section and its theoretical prediction~\cite{CMS:2019too,CMS:2021ugv}.
Theoretical uncertainties are applied to \tZq (11\%), \ZZ, \WZj, tribosons, and processes in association with a photon (10\%), and the remaining associated top quark production processes (20\%), while a 20\% uncertainty is estimated for the \WZb background~\cite{CMS:2021ugv}.
All background processes, except nonprompt leptons, are estimated using simulation where uncertainties in the normalization and modeling are considered.  The uncertainties in the integrated luminosity collected in each data-taking period (2016, 2017 and 2018) are calculated to be 1.2--2.5\%~\cite{lumipaper,CMS:LUM-17-004,CMS:LUM-18-002}, corresponding to an uncertainty of 1.6\% for the whole 2016--2018 data-taking period. Statistical uncertainties from the nonprompt-lepton background estimation, together with a 30\% normalization uncertainty in order to cover with potential residual mismodeling, are treated in a bin-wise manner. Finally, uncertainties arising from the finite size of simulated samples are also accounted for following the Barlow--Beeston method~\cite{Barlow:1993dm}. In the assessment of the systematic uncertainties, the DNNs are reevaluated on modified inputs to propagate the uncertainties to the templates.

\section{Results}
The fit provides evidence for the \tWZ process with a statistical significance of 3.4 standard deviations, for an expected one of 1.4. The signal strength is measured to be \tWZmuobs, and the cross section is found to be \tWZxsecobs. This result is two standard deviations above the SM expectations. The dominant systematic uncertainty is the \ttZ normalization with an impact on the final measurement of 18\%, constrained to the level of 10\% in the fit. Fixing the \ttZ cross section to the measured value in Ref.~\cite{CMS:2019too}, the statistical significance of the signal stays above three standard deviations.

Coming next are the uncertainties in the normalization of the backgrounds other than \ttZ and the experimental sources of uncertainty, with an impact of 12\% and 10\%, respectively. The uncertainty in the signal modeling has an impact of 5\%. The region that yields the highest sensitivity to the final measurement is the \SRa, followed by the \SRc and the \SRb. The boosted region, which includes only a small number of events, contributes with a statistical significance of less than 0.2 standard deviations. Overall, results in all categories across all data-taking periods are found to be compatible within one standard deviation.

\begin{figure}[!ht]
\centering
\ifthenelse{\boolean{cms@external}}{%
    \includegraphics[width=\linewidth]{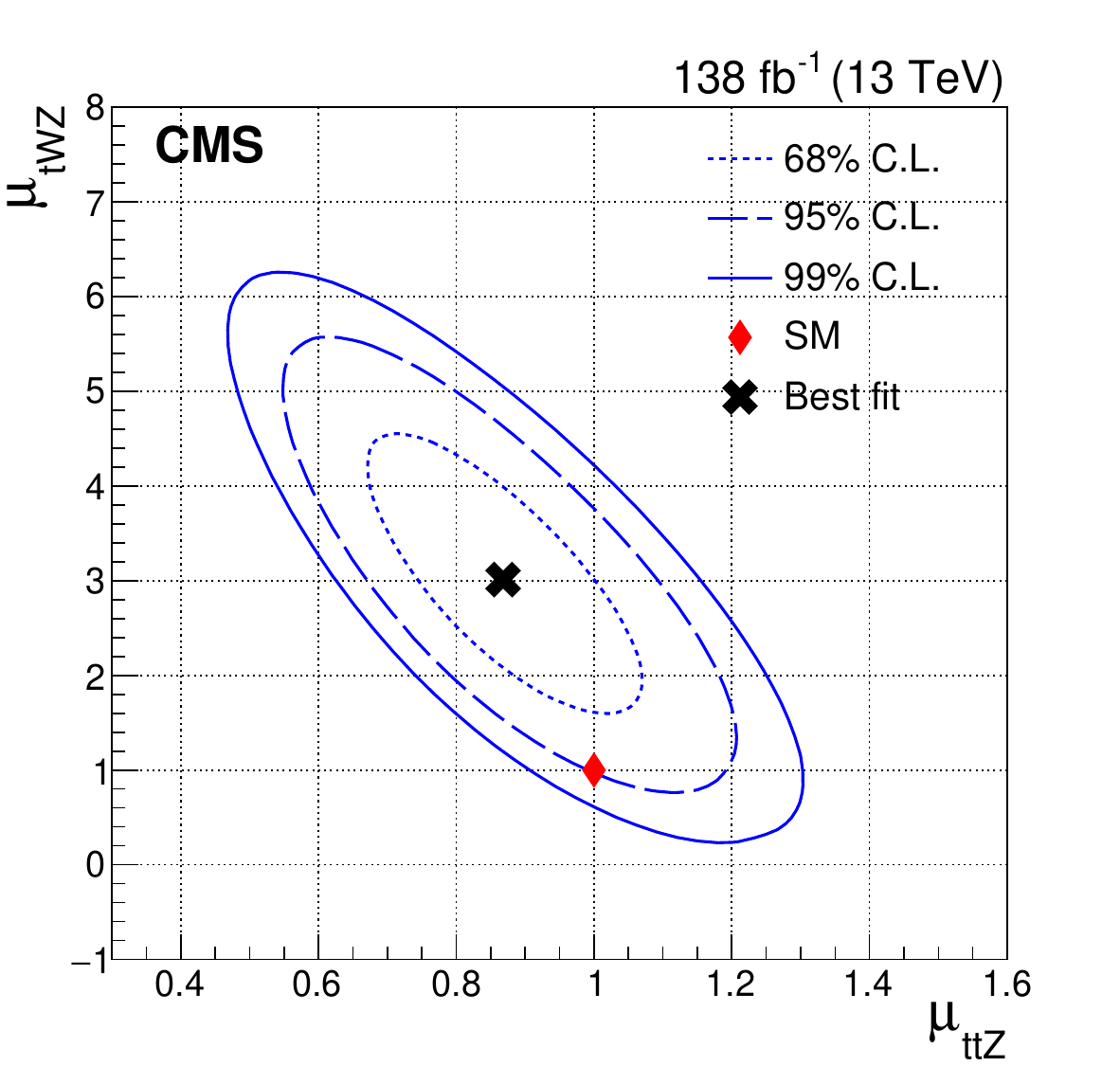}%
}{%
    \includegraphics[width=0.55\textwidth]{Figure_004.pdf}
}
\caption{Two-dimensional likelihood scan of the signal strengths of \tWZ ($\mu_{\tWZ}$) vs. \ttZ ($\mu_{\ttZ}$). The blue lines show the 68\%, 95\%, and 99\% confidence level contours. The black cross represents the best-fit value, while the red diamond the SM expectation.}
\label{fig:2Dscan}
\end{figure}

The reason for the leading impact of the \ttZ normalization uncertainty lies in the very similar nature of the \tWZ and \ttZ processes. To further investigate the interconnection, we additionally performed a simultaneous fit of the \tWZ and \ttZ signal strengths. The two-dimensional likelihood scan is displayed in Fig.~\ref{fig:2Dscan}, and shows that the two parameters are anticorrelated, with the \ttZ signal strength being compatible with the SM expectation within one standard deviation.

\section{Summary}
A measurement of the \tWZ process is performed in final states with three or four leptons, using the data collected by the CMS experiment at the LHC in 2016--2018, at a center-of-mass energy of 13\TeV, corresponding to an integrated luminosity of 138\fbinv. To assess backgrounds and establish the signal, the measurement heavily relies on multivariate techniques and the data are exploited in multiple regions and categories. The challenge of the \tWZ signal overlapping with the \ttZ background is overcome using the latest advancements in the \tWZ modeling. We find the signal to have an observed statistical significance of 3.4 standard deviations, corresponding to a measured cross section of $354\pm54\stat\pm 95\syst\unit{fb}$ that is two standard deviations above the standard model prediction. This is the first evidence of the \tWZ process.

\begin{acknowledgments}
 We congratulate our colleagues in the CERN accelerator departments for the excellent performance of the LHC and thank the technical and administrative staffs at CERN and at other CMS institutes for their contributions to the success of the CMS effort. In addition, we gratefully acknowledge the computing centers and personnel of the Worldwide LHC Computing Grid and other centers for delivering so effectively the computing infrastructure essential to our analyses. Finally, we acknowledge the enduring support for the construction and operation of the LHC, the CMS detector, and the supporting computing infrastructure provided by the following funding agencies: SC (Armenia), BMBWF and FWF (Austria); FNRS and FWO (Belgium); CNPq, CAPES, FAPERJ, FAPERGS, and FAPESP (Brazil); MES and BNSF (Bulgaria); CERN; CAS, MoST, and NSFC (China); MINCIENCIAS (Colombia); MSES and CSF (Croatia); RIF (Cyprus); SENESCYT (Ecuador); MoER, ERC PUT and ERDF (Estonia); Academy of Finland, MEC, and HIP (Finland); CEA and CNRS/IN2P3 (France); SRNSF (Georgia); BMBF, DFG, and HGF (Germany); GSRI (Greece); NKFIH (Hungary); DAE and DST (India); IPM (Iran); SFI (Ireland); INFN (Italy); MSIP and NRF (Republic of Korea); MES (Latvia); LAS (Lithuania); MOE and UM (Malaysia); BUAP, CINVESTAV, CONACYT, LNS, SEP, and UASLP-FAI (Mexico); MOS (Montenegro); MBIE (New Zealand); PAEC (Pakistan); MES and NSC (Poland); FCT (Portugal);  MESTD (Serbia); MCIN/AEI and PCTI (Spain); MOSTR (Sri Lanka); Swiss Funding Agencies (Switzerland); MST (Taipei); MHESI and NSTDA (Thailand); TUBITAK and TENMAK (Turkey); NASU (Ukraine); STFC (United Kingdom); DOE and NSF (USA).

\hyphenation{Rachada-pisek} Individuals have received support from the Marie-Curie program and the European Research Council and Horizon 2020 Grant, contract Nos.\ 675440, 724704, 752730, 758316, 765710, 824093, and COST Action CA16108 (European Union); the Leventis Foundation; the Alfred P.\ Sloan Foundation; the Alexander von Humboldt Foundation; the Science Committee, project no. 22rl-037 (Armenia); the Belgian Federal Science Policy Office; the Fonds pour la Formation \`a la Recherche dans l'Industrie et dans l'Agriculture (FRIA-Belgium); the Agentschap voor Innovatie door Wetenschap en Technologie (IWT-Belgium); the F.R.S.-FNRS and FWO (Belgium) under the ``Excellence of Science -- EOS" -- be.h project n.\ 30820817; the Beijing Municipal Science \& Technology Commission, No. Z191100007219010 and Fundamental Research Funds for the Central Universities (China); the Ministry of Education, Youth and Sports (MEYS) of the Czech Republic; the Shota Rustaveli National Science Foundation, grant FR-22-985 (Georgia); the Deutsche Forschungsgemeinschaft (DFG), under Germany's Excellence Strategy -- EXC 2121 ``Quantum Universe" -- 390833306, and under project number 400140256 - GRK2497; the Hellenic Foundation for Research and Innovation (HFRI), Project Number 2288 (Greece); the Hungarian Academy of Sciences, the New National Excellence Program - \'UNKP, the NKFIH research grants K 124845, K 124850, K 128713, K 128786, K 129058, K 131991, K 133046, K 138136, K 143460, K 143477, 2020-2.2.1-ED-2021-00181, and TKP2021-NKTA-64 (Hungary); the Council of Science and Industrial Research, India; ICSC -- National Research Center for High Performance Computing, Big Data and Quantum Computing, funded by the EU NexGeneration program (Italy); the Latvian Council of Science; the Ministry of Education and Science, project no. 2022/WK/14, and the National Science Center, contracts Opus 2021/41/B/ST2/01369 and 2021/43/B/ST2/01552 (Poland); the Funda\c{c}\~ao para a Ci\^encia e a Tecnologia, grant CEECIND/01334/2018 (Portugal); the National Priorities Research Program by Qatar National Research Fund; MCIN/AEI/10.13039/501100011033, ERDF ``a way of making Europe", and the Programa Estatal de Fomento de la Investigaci{\'o}n Cient{\'i}fica y T{\'e}cnica de Excelencia Mar\'{\i}a de Maeztu, grant MDM-2017-0765 and Programa Severo Ochoa del Principado de Asturias (Spain); the Chulalongkorn Academic into Its 2nd Century Project Advancement Project, and the National Science, Research and Innovation Fund via the Program Management Unit for Human Resources \& Institutional Development, Research and Innovation, grant B37G660013 (Thailand); the Kavli Foundation; the Nvidia Corporation; the SuperMicro Corporation; the Welch Foundation, contract C-1845; and the Weston Havens Foundation (USA).    
\end{acknowledgments}

\bibliography{auto_generated}
\cleardoublepage \appendix\section{The CMS Collaboration \label{app:collab}}\begin{sloppypar}\hyphenpenalty=5000\widowpenalty=500\clubpenalty=5000
\cmsinstitute{Yerevan Physics Institute, Yerevan, Armenia}
{\tolerance=6000
A.~Hayrapetyan, A.~Tumasyan\cmsAuthorMark{1}\cmsorcid{0009-0000-0684-6742}
\par}
\cmsinstitute{Institut f\"{u}r Hochenergiephysik, Vienna, Austria}
{\tolerance=6000
W.~Adam\cmsorcid{0000-0001-9099-4341}, J.W.~Andrejkovic, T.~Bergauer\cmsorcid{0000-0002-5786-0293}, S.~Chatterjee\cmsorcid{0000-0003-2660-0349}, K.~Damanakis\cmsorcid{0000-0001-5389-2872}, M.~Dragicevic\cmsorcid{0000-0003-1967-6783}, A.~Escalante~Del~Valle\cmsorcid{0000-0002-9702-6359}, P.S.~Hussain\cmsorcid{0000-0002-4825-5278}, M.~Jeitler\cmsAuthorMark{2}\cmsorcid{0000-0002-5141-9560}, N.~Krammer\cmsorcid{0000-0002-0548-0985}, D.~Liko\cmsorcid{0000-0002-3380-473X}, I.~Mikulec\cmsorcid{0000-0003-0385-2746}, J.~Schieck\cmsAuthorMark{2}\cmsorcid{0000-0002-1058-8093}, R.~Sch\"{o}fbeck\cmsorcid{0000-0002-2332-8784}, D.~Schwarz\cmsorcid{0000-0002-3821-7331}, M.~Sonawane\cmsorcid{0000-0003-0510-7010}, S.~Templ\cmsorcid{0000-0003-3137-5692}, W.~Waltenberger\cmsorcid{0000-0002-6215-7228}, C.-E.~Wulz\cmsAuthorMark{2}\cmsorcid{0000-0001-9226-5812}
% ...
\par}
\vskip\cmsinstskip
\dag:~Deceased\\
$^{1}$Also at Yerevan State University, Yerevan, Armenia\\
$^{2}$Also at TU Wien, Vienna, Austria\\
% ...
\end{sloppypar}
\end{document}